\documentclass[12pt]{iopart}

\usepackage{iopams}

\usepackage{epstopdf}
\usepackage{psfrag}
\usepackage{ps4pdf}

\usepackage{multirow}
\usepackage{paralist}
\usepackage{color}

\begin{document}

\newcommand{\D}{\mathrm{d}}
\newcommand{\pr}{{}^{\tiny{\mbox{пр}}}}
\newcommand{\nuint}{\nu_{\mbox{\footnotesize{инт}}}}
\newcommand{\Vreg}{V_{\mbox{\footnotesize{рег}}}}

\newcommand{\OB}[2]
{\overbrace{\hspace{#1mm}}^{ #2}}

\newcommand{\zd}
{
\scriptstyle{b-z(\tau)}\;
 \Bigg\{
}

\newcommand{\BC}{
\OB{22}{\frac{b}{c}}
}

\newcommand{\ZC}{
\OB{10}{\frac{b-z(\tau)}{c}}
}

\title{Correction due to finite speed of light in absolute gravimeters}
\author{V D Nagornyi}
\address{Axispoint, Inc., 350 Madison Avenue, New York, NY 10017, USA}
\ead{vn2@member.ams.org}
\author{Y M Zanimonskiy}
\address{Institute of Radio Astronomy, National Academy of Sciences of Ukraine, 4, Chervonopraporna St., Kharkiv, 61002, Ukraine.}
\author{Y Y Zanimonskiy}
\address{International Slavonic University, 9-A, Otakara Jarosha St., Kharkiv, 61045, Ukraine.}
\begin{abstract}
Correction due to finite speed of light is among the most inconsistent ones in absolute gravimetry. Formulas reported by different authors yield corrections scattered up to 8 $\mu$Gal with no obvious reasons. The problem, though noted before, has never been studied, and nowadays the correction is rather postulated than rigorously proven. In this paper we make an attempt to revise the subject. Like other authors, we use physical models based on signal delays and the Doppler effect, however, in implementing the models we additionally introduce two scales of time associated with moving and resting reflectors, derive a set of rules to switch between the scales, and establish the equivalence of trajectory distortions as obtained from either time delay or distance progression.  The obtained results enabled us to produce accurate correction formulas for different types of instruments, and to explain the differences in the results obtained by other authors. We found that the correction derived from the Doppler effect is accountable only for $\frac23$ of the total correction due to finite speed of light, if no signal delays are considered. Another major source of inconsistency was found in the tacit use of simplified trajectory models.
\end{abstract}
%

%
%

\newpage

\section{Introduction}
\begin{figure}[h]
\centering
\small
\input{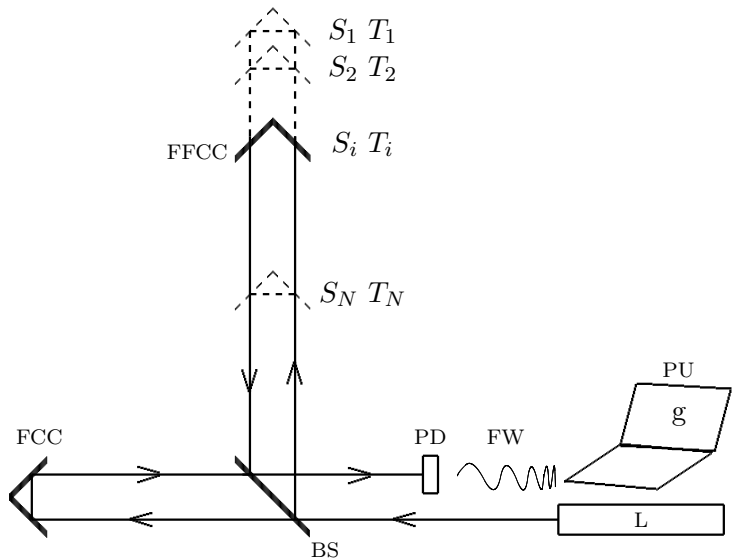}
%
%
%
%
%
%
  \caption[short title]
  {
  \quad\parbox[t]{11cm} {Functional diagram of an absolute gravimeter
  }
  }
\label{pic_c_intro}
\end{figure}

Modern absolute gravimeters measure gravity acceleration by laser tracking and subsequent analysis of the trajectory of the free falling test body. The laser beam ($L$) (fig.~\ref{pic_c_intro})
split by the mirror ($BS$) and reflected simultaneously from the fixed ($FCC$) and the free-falling ($FFCC$) corner cube qua test body, creates the interference signal converted by the photo detector ($PD$) into the fringe wave ($FW$), processed by the unit ($PU$) to obtain a set of time-distance coordinates of the falling body and calculate the gravity acceleration.
The set of coordinates $\{S_i, T_i\}$, also called ``levels'', makes up the measurement schema.  If the falling body is tracked on both upward and downward parts of the trajectory, the schema is rise-and-fall one, otherwise the schema is free-fall.
There are schemas with two, three, or four levels, as well as multi-level schemas, in which the levels may be equally spaced in time (EST), or in distance (ESD).
Although the majority of absolute gravimeters now use multi-level schemas,  in this paper we consider other schema types as well. One reason for this is that older schemas are still used, albeit rarely \cite{zhang2007}. Another reason is that theoretical results for non-multi-level schemas are often extended to the multi-level case \cite{murata1978, arnautov1983, hanada1996a}, not realizing the fact and its consequences.

Due to finite speed of light, the laser beam reflected from the free-falling corner cube, delivers the information of its position with some delay, resulting in a distortion of the measured acceleration, which therefore requires a correction.
Though the models of this phenomenon are pretty straightforward, the correction due to finite speed of light is perhaps the most controversial one in the theory of absolute gravimeters. There are more results published on this correction, than on any one else, that have later been disputed or amended by other authors.
Generally, corrections are known to depend on the measurement schema of the gravimeter. For the finite speed of light, however, the obtained corrections differ significantly even for the same schema types.
Let's examine some results published for EST schemas. Neglecting the correction sign and the test body's initial velocity\footnote{The sign of the correction depends on the position of the interferometer with respect to the test body. The initial velocity was considered not by all authors, so we neglect it for this example.}
, all authors agree that the correction is given by the formula
\begin{equation}
\label{eq_c_corr_general}
\Delta g_c = k\, \frac{g_0^2 \, T}{c},
\end{equation}
where $g_0$ is gravity acceleration, $c$ is speed of light, $T$ is the duration of the trajectory, $k$ is the disagreement factor. The values of $k$ obtained by different authors are shown in the table~\ref{tbl_k_by_authors}.
\Table{\label{tbl_k_by_authors}The values of $k$ in formula (\ref{eq_c_corr_general}) obtained by different authors.}
\br
$\0 k$ & \0 author\\
\mr
$\0 \frac23$ & \0 Murata \cite{murata1978}\\
\ms
$\0 1$ & \0 Tsubokawa T. \cite{hanada1996a}\\
\ms
$\0 \frac43$ & \0  Hanada  \cite{hanada1996a}\\
\ms
$\0 \frac32$ & \0 Kuroda \& Mio  \cite{kuroda1991}\\
\br
\endTable
As the duration of the trajectory $T$ in modern gravimeters is 0.1...0.2 s, the values of $k$ from the table
yield the corrections different as much as 5.5 $\mu$Gal
\footnote{1 $\mu$Gal = $10^{-8}$ m/s$^2$}
\footnote{The result of Hanada \cite{hanada1996a} includes a miscalculation, which, if corrected, makes his $k$ equal to $\frac13$, increasing this difference to 7.5 $\mu$Gal. The result of Hanada is discussed in chapter \ref{dm_gl_Hammond}.}.
This uncertainty exceeds significantly the precision level of modern instruments, raising the question of the ``correct'' correction.

Another way of accounting for the finite speed of light is to modify the measured time intervals $T_i$. The gravity acceleration is obtained by the model \cite{niebauer1995}:
\begin{equation}
\label{eq_c_3model}
S_i = z_0 + V_0\, \tilde {T_i} + g_0  \tilde {T_i}^2/2,
\end{equation}
where
\begin{equation}
\label{eq_c_T_cor}
\tilde {T_i} = T_i - \frac{S_i-z_0}{c}.
\end{equation}
Here $z_0$, $V_0$, and $g_0$ stand for initial coordinate, velocity, and acceleration of the test body respectively.
This model is non-linear by the parameter $z_0$, independent determination of which is rather problematic. It's also unclear if the gravity acceleration obtained with the above model can be matched to any value from the table~\ref{tbl_k_by_authors}.

In this paper we consider several approaches to the correction due to finite speed of light and suggest consistent ways for its implementation. In doing so, we uncover the reasons of the divergence of the results obtained by other authors. The paper has following structure.
In chapter \ref{dm_c_gl_obshij_podhod} we discuss the approach we use to account for disturbances in the corner cube trajectory.
In chapter \ref{dm_c_gl_vozmush} we consider several models of influence of the finite speed of light on the measured gravity value and establish connections between the models.
In chapter \ref{dm_gl_uchet_vozm} we find and investigate the corrections for different measurement schemas.
The paper is concluded by the chapter \ref{dm_gl_other_authors} where we review the results obtained by other authors.
Elements of the theory of absolute gravimeters that require additional comments for this paper, have been carried into the Appendices.
\section{Accounting for disturbances in the test body motion}
\label{dm_c_gl_obshij_podhod}
We consider the measured gravity value as weighted average of the test body acceleration $g(t)$ \cite{nagornyi1995}:

\begin{equation}
\label{eq_c_g_izm}
\overline g = \int \limits_0^T g(t)\,w(t)\,\D t,
\end{equation}
where $\overline g$ is the measured gravity,  $g(t)$ is the acceleration of the test body, $T$ is the duration of the trajectory, $w(t)$ is the gravimeter's weighting function, which  is an analog of the impulse response function of a linear system. The intrinsic property of the gravimeter's weighting function is its unit square:
\begin{equation}
\label{eq_wf_unit_square}
\int \limits_0^T w(t)\,\D t \equiv 1.
\end{equation}
If along with its normal value $g_0$, the acceleration of the test body contains a disturbance $\Delta g(t)$, the measured gravity, according to (\ref{eq_c_g_izm}), will be
\begin{equation}
\label{eq_c_g_izm2}
\overline g = \int \limits_0^T \left(g_0 + \Delta g(t)\right)\,w(t)\,\D t = g_0 + \int \limits_0^T \Delta g(t) \, w(t)\,\D t.
\end{equation}
So, if the gravimeter's weighting function is known, the additional acceleration measured due to the disturbances $\Delta g(t)$ can be found as
\begin{equation}
\label{eq_c_g_vozm}
\overline {\Delta g(t)} = \int \limits_0^T  \Delta g(t)\,w(t)\,\D t.
\end{equation}
Taken with the opposite sign, the above expression yields the correction for the considered disturbance.

If the disturbance is expanded into the power series like
\begin{equation}
\label{eq_c_dist_power_ser}
\Delta g(t) = \sum a_n\, t^n,
\end{equation}
the accounting for it is greatly simplified. In this case, the measured additional acceleration, according to (\ref{eq_c_g_vozm}), is
\begin{equation}
\label{eq_c_g_popr_via_Cn}
\overline{\Delta g(t)}
= \int \limits_0^T \left( \sum a_n\, t^n\, \right) w(t)\,\D t
= \sum a_n\, C_n,
\end{equation}
where
%
\begin{equation}
\label{eq_c_Cn_as_Int}
C_n = \int \limits_0^T t^n\, w(t)\,\D t, \;\; n = 0, 1, 2, ...
\end{equation}
is the $n$'s averaging coefficient of the gravimeter. These coefficients are tabulated for different types of measurement schemas \cite{nagornyi1995}. In addition, there is a simple formula discussed in the Appendix, that calculates averaging coefficients as function of the number and distribution of measurement levels, enabling the finesse analysis of the corrections. If no such detailed study is needed, one may employ other methods of accounting for disturbances, for example continuous least-square solution \cite{thulin1961, cook1965a, murata1978, niebauer1989}, or shifted Legendre approximation \cite{robertsson2005, robertsson2007}.
\section{Distortion of registered trajectory due to finite speed of light}
\label{dm_c_gl_vozmush}
\subsection{Distortion as signal delay}
\label{dm_gl_delay}
\subsubsection{Delays in laser beam propagation}
\begin{figure}[ht]
\centering
\small
\input{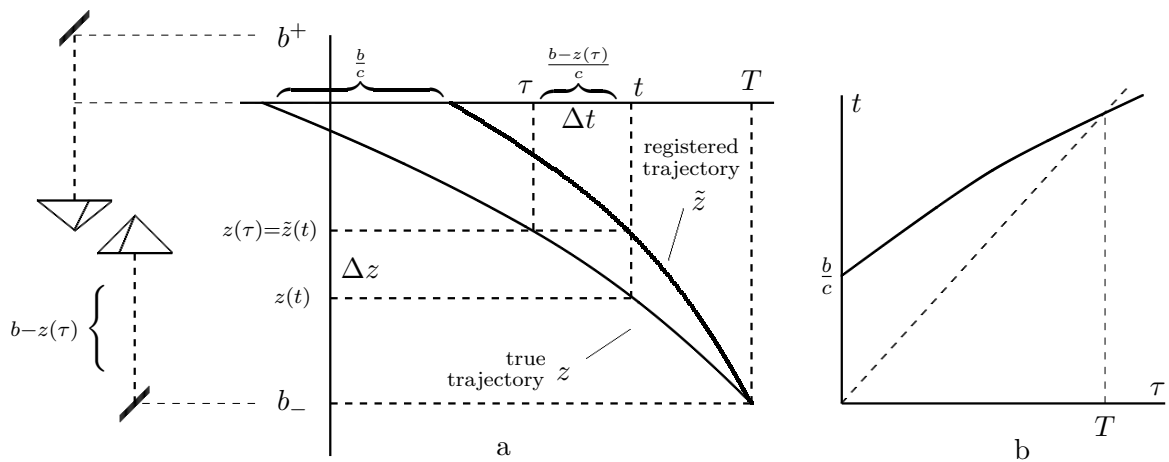}
  \caption[short title]
  {
  \quad\parbox[t]{9cm} {Distortion of the registered trajectory resulting from signal delays, viewed either as time delay $\Delta t$ or coordinate progression $\Delta z$. The case of the test body approaching the beam splitter.
    }
  }
\label{dm_C_delays}
\end{figure}

Let for some moment $\tau$ the vertical coordinate of the test body be $z(\tau)$, and the distance from its origin to the beam splitter be $b$ (fig.~\ref{dm_C_delays}а).
Then the separation from the test body to the beam splitter is  $b+z(\tau)$, if the interferometer is positioned above, or $b-z(\tau)$, if positioned below the test body. From here on, we combine both cases as $b \pm z(\tau)$, where the upper or lower sign corresponds to the beam splitter positioned above or below the test body, respectively. Due to finite speed of light, the information on the coordinate $z(\tau)$ is delivered with some delay --- in the moment $t$, for which
\begin{equation}
\label{eq_c_delay2}
t - \tau = \frac{b \pm z(\tau)}{c},
\end{equation}
where $c$ is the speed of light. So, the coordinate $z(\tau)$ is matched in data processing to the moment $t$, rather that $\tau$, causing the distortion of the trajectory. We proceed with establishing some relationships between  $t$ and $\tau$.
\subsubsection{Two scales of time}
We considered a laser beam as information carrier, traveled with some delay from the test body to the beam splitter. In this formulation, the body poses as the origin of the information, while the beam splitter poses as the information destination. The variables $\tau$ and $t$ can be considered as two scales of time. The motion of the test body is related to the scale $\tau$, whereas the registration of the motion is related to the scale $t$.

The relationship between $\tau$ and $t$ is best presented as readings of moving clock compared to the readings of the clock resting with the observer, assuming that both clocks were synchronized before the motion started. Distortion of the scale $\tau$ with respect to the scale $t$ is caused by the fact that the moving clock readings are delivered to the observer with the variable delay (\ref{eq_c_delay}). If the moving clock would also rest with the observer instead, the readings of both clocks coincided (dashed line in the fig.~\ref{dm_C_delays}b).

We use the following properties of the test body's trajectory in future analysis:
\begin{enumerate}
\item The $[\tau\,,\,t]$ interval is much shorter than the trajectory's duration $T$:
    \begin{equation}
    \label{eq_assumption_1}
    t-\tau \ll T.
    \end{equation}
    This property enables replacement of $\tau$ with $t$ in (\ref{eq_c_delay2}):
    \begin{equation}
    \label{eq_c_delay}
    t - \tau = \frac{b \pm z(t)}{c},
    \end{equation}
\item The distance traveled by the test body during the $[\tau\,,\,t]$ interval is much shorter than the distance to the beam splitter:
    \begin{equation}
    \label{eq_assumption_2}
    |z(t) - z(\tau)| \ll b.
    \end{equation}
\item Change in the body's velocity during the $[\tau\,,\,t]$ interval is minuscule compared to the change over the entire trajectory:
    \begin{equation}
    \label{eq_assumption_3}
    \dot z(\tau) - \dot z(t) \approx 0.
    \end{equation}
\end{enumerate}
Simplifications coming from these properties are based on the fact that the timescales $t$ and $\tau$ differ in terms $\sim O(c^{-1})$ (\ref{eq_c_delay2}), so interchanging the scales in expressions having another $c$ in denominator creates difference terms of only $\sim O(c^{-2})$, which are negligible as the test body's velocity is diminutive compared to the speed of light.

\subsubsection{The distortion as obtained from signal delay}
\label{gl_delay}
Two ways are possible to deduce the trajectory distortion from signal delays.
\paragraph{Adjustment of time.}
We can proceed from the fact that the coordinate registered at the moment $t$ corresponds to the true coordinate at the earlier moment $\tau$:
\begin{equation}
\label{eq_c_z_t-tau}
\tilde z(t) = z(\tau).
\end{equation}
Expressing $\tau$ via $t$ using (\ref{eq_c_delay}), we get the following formula for the registered trajectory:
\begin{equation}
\label{eq_c_z_reg_time}
\tilde z(t) = z\left(t-\frac{b \pm z(t)}{c}\right).
\end{equation}
\paragraph{Adjustment of coordinate.}
While the laser beam is traveling to the beam splitter, the test body keeps on moving along its path, so at the moment the beam reaches  the beam splitter the body's true coordinate has progressed a bit compared to the registered one:
\begin{equation}
\label{eq_c_z_reg}
z(t) = \tilde z(t) + \Delta z.
\end{equation}
As, according to  (\ref{eq_assumption_3}) the body's velocity changes insignificantly over the path $\Delta z$, we can use
\begin{equation}
\label{eq_c_Delta_z}
\Delta z = \dot z(t) (t - \tau).
\end{equation}
Substituting this expression into (\ref{eq_c_z_reg}) and applying (\ref{eq_c_delay}), we find
\begin{equation}
\label{eq_c_z_reg_path}
\tilde z(t) = z(t) \left(1 \mp \frac{\dot z(t)}{c} \right)
- b \frac{\dot z(t)}{c} .
\end{equation}

The above formula also follows from (\ref{eq_c_z_reg_time}) by Taylor expansion
, which makes both formulas equivalent for future analysis. The registered acceleration we find from (\ref{eq_c_z_reg_path}) as
\begin{equation}
\label{eq_c_delay_distortion}
g(t) = \label{eq_c_reg_USP}
\ddot {\tilde z}(t) = \ddot z(t)  \mp \frac {3\dot z(t)\ddot z(t) }{c}
- \frac{\tdot z(t)}{c} (b \pm z(t)).
\end{equation}
For uniformly accelerated motion
\begin{equation}
\label{eq_uniform_acc_in_tau}
z(t) = z_0 + V_0\,t + g_0\,t^2/2
\end{equation}
the registered acceleration, according to (\ref{eq_c_reg_USP}), is
\begin{equation}
\label{eq_Delaa_g_c}
g(t) =
g_0 \mp \frac{3\,g_0}{c}(V_0 + g_0\,t).
\end{equation}
\emph{Remark}. Equivalence of the formulas  (\ref{eq_c_z_reg_time}) and (\ref{eq_c_z_reg_path}) allows the trajectory distortion to be obtained  as either time delay or coordinate progression. Combining both approaches, however, will cause incorrect doubling of the distortion.
\subsection{Doppler distortion of the registered trajectory}
\label{section_on_Doppler}
The reasoning we used to get the additional term of the acceleration in (\ref{eq_Delaa_g_c}) was based on abstract   considerations regarding signal delays. We did not rely on any specifics of a particular signal type, like passive or active, pulse or continuous, etc. Neither did we specify how the information is coded with the signal. Historically, however, analysis of the correction due to finite speed of light relied on the implemented physical phenomenon of interference of direct and reflected laser beams, manifesting itself as Doppler effect.
The frequency of the light wave reflected from an object moving with the velocity $V(\tau)$ after experiencing two Doppler shifts is given by {\cite{pauli1981}}
\begin{equation}
\label{dm_v_5}
\nu(\tau) = \nu_0 \; \frac{c \mp V(\tau)}{c \pm V(\tau)},
\end{equation}
where $\nu_0$ is the frequency before reflection. The frequency of the interference signal, neglecting terms $\sim O(V/c)^2$, will then be
\begin{equation}
\label{dm_v_interf}
\tilde \nu(\tau) = \mp (\nu(\tau) - \nu_0) = 2\,\nu_0 \; \frac{V(\tau)}{c \pm V(\tau)} = \frac{2\nu_0}{c} \; \frac{ V(\tau)}{1 \pm \frac{V(\tau)}{c}} =
\frac{2\nu_0}{c} V(\tau) \left(  1 \mp \frac{V(\tau)}{c} \right) .
\end{equation}
%
The velocity of the moving reflector can be found as
\begin{equation}
\label{dm_v_reg_Doppler}
\tilde V (\tau) = \frac{\lambda}{2} \tilde \nu(\tau) = V(\tau) \mp \frac{V^2(\tau)}{c},
\end{equation}
because $\lambda \nu_0 = c$, where $\lambda$ is the laser wavelength. For uniformly accelerated motion $V(\tau) = V_0 + g_0\;\tau$ we get
\begin{equation}
\label{dm_v_6000}
\tilde V (\tau) = V_0 + g_0\;\tau \mp \frac{(V_0 + g_0\;\tau)^2}{c}.
\end{equation}
Acceleration found as $\D \tilde V /\D \tau$ would be
\begin{equation}
\label{eq_vozm_Doppler_only}
\tilde g (\tau) = g_0 \mp \frac{2\,g_0}{c} (V_0 + g_0 \tau).
\end{equation}
Additional acceleration term of this formula differs from that obtained earlier in (\ref{eq_Delaa_g_c}). The reason is that we implicitly assumed the interference (\ref{dm_v_interf}) to take place at the moment of reflection, while it's actually taking place with the delay defined by (\ref{eq_c_delay}). To account for the delay, we need to convert the scale $\tau$ into the scale $t$. Substituting (\ref{eq_c_delay}) into (\ref{dm_v_6000}), we get:
\begin{equation}
\label{eq_uniform_V_distorted}
\tilde V(t) = V_0 + g_0\;\left(t - \frac{b \pm V_0\,t\pm g_0\,t^2/2}{c}\right) \mp \frac{\left(V_0 + g_0\;\left(t - \frac{b \pm V_0\,t \pm g_0\,t^2/2}{c}\right)\right)^2}{c},
\end{equation}
As before, taking the derivative and dropping components $\sim O(V/c)^2$ or less, we get
\begin{equation}
\label{eq_uniform_g_w_additional_c_term}
g(t) = g_0 \mp \frac{3\,g_0}{c}(V_0 + g_0\,t).
\end{equation}
\subsection{``Correction due to the Doppler effect'' or ``correction due to finite speed of light?''}
\label{dm_c_gl_doppler_or_delay}
We are ready to highlight a misconception responsible for a significant part of discrepancies in the results, obtained by different authors. We first have to admit that both terms in the title of the current chapter are misnomers. Physical phenomena giving names to corrections are usually of a secondary nature, and their influence on the measurement result is relatively small. In this view, the Doppler effect is a primary phenomenon, making interferometric measurement of gravity at all possible. The finiteness of the speed of light, in turn, is the reason for the Doppler effect to exist, so it's also a primary phenomenon. More accurate would be the correction terms like  ``non-linearity of the Doppler effect'' and ``non-uniform delays of the light signal''. It's with these clarifications in mind, that we still use the old, historically settled terms in this paper.

For a long time both terms were used as synonyms. In fact, the correction due to the Doppler effect accounts only for the $\frac23$ of the total correction due to finite speed of light, if, as shown above,
no signal delay is taken into consideration.

If the Doppler effect is used for the correction, the ``missing'' $\frac13$ of the total value can be ``recovered'' by rendering the reflectors's velocity into the scale $t$. Indeed, if the actual velocity is
\begin{equation}
\label{eq_uniform_vel_in_tau}
V(\tau) = V_0 + g_0\,\tau,
\end{equation}
the observed velocity, because of (\ref{eq_c_delay}), will be
\begin{equation}
\label{eq_uniform_vel_in_t}
V(t) = V_0 + g_0\,\left(t - \frac{b \pm V_0\,t \pm g_0\,t^2/2}{c}\right),
\end{equation}
which is equivalent to the observed acceleration found as $\D V /\D t$:
\begin{equation}
\label{eq_uniform_g_distorted_by_delay_in_V}
g(t) = g_0 \mp \frac{g_0}{c}(V_0 + g_0\,t),
\end{equation}
representing exactly $\frac13$
of the additional component in (\ref{eq_uniform_g_w_additional_c_term}).
\section{Compensating distortions from the finite speed of light}
\label{dm_gl_uchet_vozm}
\subsection{Correcting the measured gravity}
By averaging the distorted acceleration (\ref{eq_Delaa_g_c}) with the weighting function (\ref{eq_c_g_vozm}), we find, using (\ref{eq_c_Cn_as_Int}), that the additional value measured by the gravimeter due to finite speed of light is
\begin{equation}
\label{eq_delta_g_c}
\overline {\Delta g_c} = \mp \frac{3\,g_0}{c}(V_0 + g_0\,C_1).
\end{equation}
Taken with the opposite sign, the above expression 
yields the correction due to finite speed of light. We now find the correction for different measurement schemas.
\subsubsection{Two-level schema}
Substituting (\ref{eq_c_C1_2_level}) into (\ref{eq_delta_g_c}) and  allowing the only sensible value of $V_0=0$ for this schema, we get
\begin{equation}
\label{eq_c_corr_2_level}
\Delta g_c = \pm \frac{g_0^2\,T}{c}.
\end{equation}
\subsubsection{Three-level schema}
Substituting (\ref{eq_c_C1_3_level}) into (\ref{eq_delta_g_c}), we get
\begin{equation}
\label{eq_c_corr_3_level}
\Delta g_c = \pm \frac{g_0}{c}\left(g_0\,(T_1+T_2) + 3\,V_0 \right).
\end{equation}
\subsubsection{Four-level schema}
\label{dm_gl_4_levels}
Substituting (\ref{eq_c_C1_4_level}) into (\ref{eq_delta_g_c}), we get
\begin{equation}
\label{eq_c_corr_eq_T}
\Delta g_c = \pm  \frac{g_0}{c}
\left(
g_0\,\frac{T_3^2 + T_2^2 - T_1^2 + T_2 T_3}{T_3+T_2-T_1}
+ 3\,V_0
\right).
\end{equation}
If $T_3-T_2=T_1,\;\;T_2-T_1=\tau$, the correction will be
\begin{equation}
\label{eq_c_arnautov_4_lev_}
\Delta g_c = \pm \frac{3 g_0}{2c} \left( g_0\,(2 T_1 +\tau) + 2\,V_0  \right).
\end{equation}
\subsubsection{EST schemas}
For levels equally spaced in time, the correction due to finite speed of light, according to (\ref{eq_C1_T}), does not depend on the number of levels and equals
\begin{equation}
\label{eq_c_corr_eq_T}
\Delta g_c = \pm \frac{3\,g_0}{c}\left(V_0 + \frac12 \, g_0\,T \right).
\end{equation}
\subsubsection{ESD schemas}
\label{c_po_puti}
\paragraph{Accurate calculation of the correction.}
Because the averaging coefficient $C_1$ in ESD case depends on the number of levels, the accurate correction value can be obtained by calculating the coefficient with the formula (\ref{eq_Ti_spaced}), and substituting it into (\ref{eq_delta_g_c}), resulting in
\begin{equation}
\label{eq_c_corr_accurate}
\Delta g_c = \pm \frac{3\,g_0}{c}\left(V_0 + \frac{g_0}{6}
G\left(T_i \, , \, T_i^3\right)
\right),
\end{equation}
where the vector $T_i$, $i=1...N$ is calculated using (\ref{eq_c_C1}).
\paragraph{Case $N \rightarrow \infty$.} For infinite number of levels, the calculations can be simplified using the $C_1$ coefficient given by (\ref{c_ESD_C1_VO}), which yields the following correction formula:
\begin{equation}
\label{eq_c_corr_ESD_V0}
\Delta g_c = \pm \frac{3\,g_0}{c}\left(V_0 +
g_0 \frac{T\,\left( 4\,g_0^3\,T^3 +
      45\,g_0^2\,T^2\,V_0 +
      108\,g_0\,T\,V_0^2 + 70\,V_0^3
      \right) }{7\,
    \left( g_0^3\,T^3 +
      12\,g_0^2\,T^2\,V_0 +
      30\,g_0\,T\,V_0^2 + 20\,V_0^3
      \right) }
\right).
\end{equation}
\paragraph{Case $N \rightarrow \infty$ and $V_0=0$ while calculating $C_1$.} Simplified formula (\ref{c_ESD_C1_VO_0}) for $C_1$ leads to the following correction:
\begin{equation}
\label{eq_c_corr_ESD_V0_0}
\Delta g_c = \pm \frac{3\,g_0}{c}\left(V_0 + \frac47 \, g_0\,T
\right),
\end{equation}
also given by Kuroda \& Mio in \cite{kuroda1991}. It must be stressed here, that even though the above formula includes the term $V_0$, it does not completely account for the effects of initial velocity, because it was assumed to be zero when obtaining the averaging coefficient $\frac47 T$.
To see how accurate the corrections given by the simplified formulas (\ref{eq_c_corr_ESD_V0_0}) and (\ref{eq_c_corr_ESD_V0}) are, in the table~\ref{tbl_c_ESD_corrections} we find the corrections for some real instruments that use ESD schema.
\Table{\label{tbl_c_ESD_corrections} Corrections due to finite speed of light for some instruments using ESD schemas:\\
(\ref{eq_c_corr_accurate}) is the accurate correction.\\
(\ref{eq_c_corr_ESD_V0}) is the approximation for $N \rightarrow \infty$.\\
(\ref{eq_c_corr_ESD_V0_0}) is the Kuroda \& Mio's approximation \cite{kuroda1991}.
%
%
}
\br
& &\centre{3}{schema parameters}&\centre{3}{$\Delta g_c$, $\mu$Gal, by formula:}\\
\ns
\ns
gravimeter & setup & \crule{3} & \crule{3}\\
 & & $V_0$, $\frac{m}{s}$ & $T$, s & \0\0\0$N$ & (\ref{eq_c_corr_ESD_V0_0}) & (\ref{eq_c_corr_ESD_V0}) & (\ref{eq_c_corr_accurate})\\
\mr

JILA & \cite{zumberge1982} & $0.30$ & $0.20$ & \0\0\0\0\0$45$ & $-13.94$ & $-13.35$ & $-13.11$\\
JILAg & \cite{klopping1991} & $0.20$ & $0.19$ & \0\0\0\0$170$ & $-12.41$ & $-11.95$ & $-11.87$\\
FG5 & \cite{niebauer1995} & $0.40$ & $0.17$ & \0\0\0\0$700$ & $-13.27$ & $-12.66$ & $-12.65$\\
A10 & \cite{schmerge2006} & $0.30$ & $0.09$ & \0\0\0\0$200$ & $-\0 7.89$ & $-\0 7.52$ & $-\0 7.51$\\
MPG-1 & \cite{rothleitner2009} & $0.32$ & $0.20$ & $1635000$ & $-14.14$ & $-13.53$ & $-13.53$\\
MPG-2 & \cite{svitlov2010a} & $0.32$ & $0.16$ & $1100000$ & $-11.94$ & $-11.40$ & $-11.40$\\
\br
\endTable
As follows from the table,
the magnitude of the correction (\ref{eq_c_corr_ESD_V0_0}) by Kuroda \& Mio \cite{kuroda1991} can for real instruments exceed the accurate value up to 1 $\mu$Gal. On the other hand, the approximation (\ref{eq_c_corr_ESD_V0}) for $N \rightarrow \infty$ produces  corrections accurate up to the 0.01 $\mu$Gal, if the number of levels is at least several hundreds.
\subsubsection{Symmetric rise-and-fall schema}
Because of the (\ref{eq_c_C1_symm}), the correction due to finite speed of light is negligible for the symmetric rise-and-fall schemas.
\subsection{Refining the trajectory model}
\label{dm_gl_c_utochn_trajekt}
Due to the signal delays, the model of the uniformly accelerated motion%
\begin{equation}
\label{eq_parabola_disc}
S_i = z_0 + V_0\,T_i + g_0\,T_i^2/2 + \epsilon_i,
\end{equation}
where\\
$T_i$
are the measured time intervals,\\
$S_i$ are corresponding measured distances,\\
$\epsilon_i$ are the measurement errors,\\
is statistically inadequate, if the time intervals are related to the beam splitter. The least squares estimate (\ref{eq_c_G3}) of the parameter $g_0$ in this case has a bias defined by (\ref{eq_delta_g_c}).
To bring the model adequate, the time intervals without delays should be used, i.e. the time intervals related to the reflector, not the beam splitter.
This transition is inverse to what we did in the chapter~\ref{gl_delay}, to find the correction (statistically: bias). The refined model is
\begin{equation}
\label{eq_parabola_disc}
S_i = z_0 + V_0\,\tilde{T_i} + g_0\,\tilde{T_i}^2/2 + \epsilon_i,
\end{equation}
where
\begin{equation}
\label{eq_Ti_corr}
\tilde{T_i} = T_i + \frac{b \pm S_i}{c}.
\end{equation}
The property (\ref{eq_assumption_2}) enables us to use the measured distances instead of the real ones.
Time intervals corrected as above will result in unbiased least squares estimate of the parameter $g_0$, so no further correction is needed. The only caveat in this approach is to estimate the parameter $b$.
One approach \cite{niebauer1995} would be to use $z_0$ of (\ref{eq_parabola_disc}) as $b$.
Unfortunately, there is no way to arrange data acquisition and processing in absolute gravimeters so that parameter $z_0$ would represent the  distance from the beam splitter to the initial position of the reflector
. Moreover, identifying $b$ with $z_0$ would render the model (\ref{eq_parabola_disc}) non-linear with respect to $z_0$.
Another approach, suggested in \cite{rothleitner2008}, is to put $b=0$. The possibility to nullify $b$, as well as to assign it any arbitrary value follows from insensitivity of the estimate (\ref{eq_c_G3}) to any constant additive of time \cite{nagornyi1993eng}:
\begin{equation}
\label{eq_GE_Ti_plus_A}
G(T_i + A \, , \, S_i) = G(T_i \, , \, S_i),
\end{equation}
where $A$ is some constant, which in our case equals $b/c$. The formula for the corrected time intervals would then be
\begin{equation}
\label{eq_Ti_corr_via_Zi_no_b}
\tilde{T_i} = T_i \pm \frac{S_i}{c}.
\end{equation}
The property (\ref{eq_GE_Ti_plus_A}) also holds true for the tree- and four-level models (\ref{eq_c_g_izm_G_3_level}) and (\ref{eq_c_g_izm_G_4_level}), so the simplification (\ref{eq_Ti_corr_via_Zi_no_b}) can be used for these models as well, but not for the models (\ref{eq_c_g_izm_G_2_level}) or (\ref{eq_c_G2}).
\section{Corrections due to finite speed of light obtained by other authors}
\label{dm_gl_other_authors}
\subsection{Hammond and Faller
\cite{hammond1971}
}
\label{Hammond_Faller}
The correction reported by the authors for the three-level schema is
\begin{equation}
\label{eq_Hammond_Faller_c_corr}
\Delta g_c = -\frac{g_0}{c} \left( \frac{4}{3}\,g_0\,(T_1+T_2) + 2\,V_0  \right).
\end{equation}
The discrepancy with the result (\ref{eq_c_corr_3_level}) is partially explained by identifying the corrections due to the Doppler effect and due to finite speed of light, as discussed in the chapter~\ref{section_on_Doppler}. In that case the coefficient 
$\frac43$ should be $\frac23$.
\subsection{Arnautov \etal
\cite{arnautov1972eng, arnautov1974, arnautov1979, arnautov1983}
}
Using the phase progression, in the paper \cite{arnautov1972eng} the authors assessed the trajectory distortion similar to (\ref{eq_c_corr_3_level}) and found the correction for the three-level schema same as (\ref{eq_c_z_reg_path}).
For the specific case of four-level schema we considered in  chapter ~\ref{dm_gl_4_levels}, the correction obtained by the authors \cite{arnautov1972eng, arnautov1974} is thrice less than in (\ref{eq_c_arnautov_4_lev_}).

In the paper \cite{arnautov1983}, the authors applied the formula (\ref{eq_c_G2}) to the four-level schema. The correction given by the authors corresponds to the coefficient $C_1$=$(T_1+T_2+T_3)/3$. According to our estimates, the $C_1$ in  this case is defined by (\ref{eq_c_C1_4_multilevel}).
\subsection{Murata
\cite{murata1978}
}
The paper \cite{murata1978} deals with the EST schema, for which the correction we found is (\ref{eq_c_corr_eq_T}). The paper reports the following formula for the correction
\begin{equation}
\label{eq_Murata_c_corr}
\Delta g_c = \frac{2}{3} \, \frac{g_0^2}{c}\,T.
\end{equation}
To get the correction, the Doppler effect (only) was applied to the two-level schema, making the result equal $\frac23$ of the correction for such a schema (\ref{eq_c_corr_2_level}).
%
%
\subsection{Zumberge
\cite{zumberge1981}
}
The gravimeter described in \cite{zumberge1981} implements the ESD schema, but the applied correction agrees with that for the EST schema (\ref{eq_c_corr_eq_T}). To get the correction, the author averages acceleration like we do in (\ref{eq_c_g_izm2}), but with no weighting function. This averaging actually corresponds to the uniform weighting function $w(t)\equiv \frac{1}{T}$, having the first averaging coefficient
\begin{equation}
\label{eq_c_C1_uniform}
C_1 = \frac{1}{T}\int \limits_0^T t \, \D t = \frac{T}{2},
\end{equation}
which coincidentally corresponds to the value for the EST schema (\ref{eq_C1_T}), thus explaining the obtained result.\footnote{This coincidence can also be explained by symmetry of the uniform function --- see formula {\ref{c_wf_sym}} }
\subsection{Kuroda and Mio
\cite{kuroda1991}
}
Considering both signal delay and Doppler effect, the authors found the interference phase, and integrated it to get the distorted trajectory. For the three-level schema (\ref{eq_c_g_izm_G_3_level}) the correction was found to be
\begin{equation}
\label{eq_Hammond_Faller_c_corr_kuroda_mio}
\Delta g_c = -\frac{g_0}{c} \left( g_0\,(T_1+T_2) + 2\,V_0  \right).
\end{equation}
For the similar schema described in the paper \cite{yongyuan1982}, the correction was determined as
\begin{equation}
\label{eq_yong_c_corr_kuroda_mio}
\Delta g_c = -\frac{g_0}{c} \left( g_0\,(t_1+t_2+t_3) + 3\,V_0  \right),
\end{equation}
where $t_1$ and $t_2$ are the start and the end of the $T_1$ interval, $t_3$ is the end of the $T_2$ interval (fig.~\ref{pic: 2-3-4_levels}~b). In the first formula, the coefficient of $V_0$ should be 3. The second formula is valid when $t_1$ coincides with the time origin.

Asymptotic value of the correction the authors obtained for the EST schema is the same as in (\ref{eq_c_corr_eq_T}). For the ESD schema, the authors got
\begin{equation}
\label{eq_c_corr_eq_S_lim}
\Delta g_c =  \pm \frac{3\,g_0}{c}(V_0 + \frac{4}{7} \, g_0\,T).
\end{equation}
The coefficient $\frac{4}{7}$ of the above formula corresponds to zero initial velocity and infinite number of levels.  For real cases, as shown in the chapter~\ref{c_po_puti}, the above correction may up to 1 $\mu$Gal exceed in absolute value the accurate one.
%
%
\subsection{Hanada
\cite{hanada1996a}
}
\label{dm_gl_Hammond}
The paper evaluates the distortion of the registered trajectory based on signal delays. For a small increment of the observed time $\Delta t$, the corresponding distance $\Delta z$ is found. The total length and duration of the trajectory are determined by integration, and the acceleration is found using two-level schema (\ref{eq_c_g_izm_G_2_level}). One should expect this approach to yield the correction like (\ref{eq_c_corr_2_level}). Indeed, the increment of the real time corresponding to the $\Delta t$ increment of the observed time,  can be found based on (\ref{eq_c_delay}) as\footnote{Following the author, we assume the trajectory of the test body to be $z(\tau)=g_0 \tau^2/2.$}
\begin{equation}
\label{eq_Hanada_dt}
\Delta \tau = \left( 1+\frac{g_0 \, t}{c} \right) \Delta t.
\end{equation}
The increment of the distance traveled in time  $\Delta \tau$ is
\begin{equation}
\label{eq_Hanada_dz}
\Delta z = g_0 \, \tau \, \Delta \tau.
\end{equation}
Expressing $\tau$ in terms of $t$ using  (\ref{eq_c_delay}) and substituting $\Delta \tau$ from (\ref{eq_Hanada_dt}), we find the total distance registered in time  $T$:
\begin{equation}
\label{eq_Hanada_z}
\overline S = \int \limits_{0}^{T} g_0 \, \tau \, \Delta \tau =
g_0 \int \limits_{0}^{T} \left( t + \frac{g_0 \, t^2}{2c} \right)\left( 1+\frac{g_0 \, t}{c} \right) \D t
=
\frac{g_0\, T^2}{2} + \frac{g_0^2\, T^3}{2c}
.
\end{equation}
The registered time is
\begin{equation}
\label{eq_Hanada_T}
\overline T = \int \limits_{0}^{T} \Delta t = \int \limits_{0}^{T} \D t = T,
\end{equation}
leading to the following registered acceleration
\begin{equation}
\label{eq_Hanada_g}
\overline g =\frac{2\overline S}{\overline T^2}
= g_0 + \frac{g_0^2\, T}{c}.
\end{equation}
We've just very likely followed T. Tsubokawa to his unit value for the disagreement coefficient $k$ in (\ref{eq_c_corr_general}), as cited by Hanada on pages 99 and 134 of his paper. However, the value obtained by Hanada himself is $\frac43$.  The discrepancy is the result of the following assumptions.
\begin{enumerate}
\item The information-bearing signal is considered propagating from the beam splitter to the reflector, rather than vice versa, causing the opposite sign of the time distortion. The formula (3.6.22) of the paper (corresponding to our formula (\ref{eq_Hanada_dt})) should have the opposite sign.
\item Expressing the registered distance via the body's velocity, the time distortion is accounted for only in the time increment, but not in the time itself. The formula (3.6.24) of the paper treats the distance increment as $g_0 \, t \, \Delta \tau$, rather than $g_0 \, \tau \, \Delta \tau$, like in (\ref{eq_Hanada_dz})
\item The paper implements double accounting of signal delays, as mentioned at the end of the chapter \ref{gl_delay}. The paper's formula (3.6.23) treats the registered time as an integral of $\Delta \tau$, rather than that of $\Delta t$, like in (\ref{eq_Hanada_T})
\item The factor 2 in the paper's formula (3.6.22) should be 1.
\end{enumerate}
While every assumption changes the result in its own way, their cumulative effect can be found with the formula
\begin{equation}
\label{eq_Hanada_all_factors}
\overline g = 2g_0 \, \frac
{
\int \limits_{0}^{T} \left(t + P\,D\,\frac{g_0 \, t^2}{2c} \right)
\left( 1 + D\,\frac{g_0 \, t}{c} \right) \D t
}
{
\left(\int \limits_{0}^{T} \left(1 + P\,Q\,D\,\frac{g_0 \, t}{c} \right)  \D t \right)^2
}
,
\end{equation}
being just the two-level schema formula $\overline g =2S/T^2$, in which the registered distance and time are expressed through the formulas (\ref{eq_Hanada_z}) and (\ref{eq_Hanada_T}) with the option to include or exclude the above assumptions. For example, the sign of time distortion (assumption (i)) is modeled by the factor $D$ assuming values: 1 or -1 (see table~\ref{tbl_Hadada_c_factors}).
The above expression
is equivalent to
\begin{equation}
\label{eq_Hanada_all_factors_result}
\overline g = g_0
\left(1+\frac{g_0\,T}{c}\,
\left(
\frac13\,P\,D + \frac23\,D - R\,Q\,D
\right)
\right)
.
\end{equation}
So, the correction is influenced by the assumptions in the following way
\begin{equation}
\label{eq_Hanada_c_via_k}
\Delta g_c = - k\,\frac{g_0^2\,T}{c},
\end{equation}
where
\begin{equation}
\label{eq_Hanada_c__k}
k = \frac13\,P\,D + \frac23\,D - R\,Q\,D.
\end{equation}
The table \ref{tbl_Hadada_c_factors} shows possible values of $D$, $P$,$Q$, $R$ , and the resulting values of $k$ in formula (\ref{eq_Hanada_c_via_k}), which cover the range from $-\frac43$ to $\frac43$ with the increment of $\frac13$.
\Table{\label{tbl_Hadada_c_factors}The factors of the formula (\ref{eq_Hanada_all_factors_result}), emulating the influence of different assumptions made in the paper
\cite{hanada1996a}
on the correction due to finite speed of light.}
\br
\centre{4}{assumptions}&\centre{4}{factors}\\
\ns
\crule{4} &\crule{4} & $k$\\
(i) & (ii) & (iii) & (iv) & $D$ & $P$ & $Q$ & $R$ &  \\
\mr

$-$ & $-$ & $-$ & $-$ & 1 & 1 & 0 & 1 & $1^{\rm a}$   \\
\ms
$-$ & $-$ & $-$ & $\checkmark$ & 1 & 1 & 0 & 2 & $1$  \\
\ms
$-$ & $-$ & $\checkmark$ & $-$ & 1 & 1 & 1 & 1 & $0$  \\
\ms
$-$ & $-$ & $\checkmark$ & $\checkmark$ & 1 & 1 & 1 & 2 & \-1  \\
\ms
$-$ & $\checkmark$ & $-$ & $-$ & 1 & 0 & 0 & 1 & $\frac23$  \\
\ms
$-$ & $\checkmark$ & $-$ & $\checkmark$ & 1 & 0 & 0 & 2 & $\frac23$ \\
\ms
$-$ & $\checkmark$ & $\checkmark$ & $-$ & 1 & 0 & 1 & 1 & \-$\frac13$ \\
\ms
$-$ & $\checkmark$ & $\checkmark$ & $\checkmark$ & 1 & 0 & 1 & 2 & \-$\frac43$  \\
\ms
$\checkmark$ & $-$ & $-$ & $-$ & \-1 & 1 & 0 & 1 & \-1  \\
\ms
$\checkmark$ & $-$ & $-$ & $\checkmark$ & \-1 & 1 & 0 & 2 & \-1  \\
\ms
$\checkmark$ & $-$ & $\checkmark$ & $-$ & \-1 & 1 & 1 & 1 & $0$  \\
\ms
$\checkmark$ & $-$ & $\checkmark$ & $\checkmark$ & \-1 & 1 & 1 & 2 & $1$  \\
\ms
$\checkmark$ & $\checkmark$ & $-$ & $-$ & \-1 & 0 & 0 & 1 & \-$\frac23$  \\
\ms
$\checkmark$ & $\checkmark$ & $-$ & $\checkmark$ & \-1 & 0 & 0 & 2 & \-$\frac23$  \\
\ms
$\checkmark$ & $\checkmark$ & $\checkmark$ & $-$ & \-1 & 0 & 1 & 1 & $\frac13^{\rm b}$ \\
\ms
$\checkmark$ & $\checkmark$ & $\checkmark$ & $\checkmark$ & \-1 & 0 & 1 & 2 & $\frac43^{\rm c}$ \\
\br
\end{tabular}
\item[] $^{\rm a}$ The result of T. Tsubokawa.
\item[] $^{\rm b}$ The result of Hanada, no evaluation error.
\item[] $^{\rm c}$ The result of Hanada.
\end{indented}
\end{table}
\subsection{Niebauer \etal
\cite{niebauer1995}
}
The instrument described in the paper \cite{niebauer1995} implements refining of the trajectory approach, which, according to the chapter~\ref{dm_gl_c_utochn_trajekt}, gives accurate correction value. Discussing the correction, the paper presents its approximate value as $-\frac32 \frac{g_0^2 T}{c}$\footnote{We have fixed the obvious typo in the text.}, which corresponds to the $V_0=0$ case of the EST schema (\ref{eq_c_corr_eq_T}). As the instrument actually implements the ESD schema, more accurate approximation is given by
the formula (\ref{eq_c_corr_ESD_V0_0}), i.e. $-\frac{12}{7} \frac{g_0^2 T}{c}$. Two approximations differing 1.4 $\mu$Gal suggest some degree of uncertainty in the knowledge about the correction \cite{kacker2007}, which probably should have been reflected in the error budget, as it includes entries as low as 0.1 $\mu$Gal.
\subsection{Nagornyi
\cite{nagornyi1995}
}
The approach suggested by Kuroda \& Mio in \cite{kuroda1991} was followed with the weighting functions treatment, as in chapter \ref{dm_c_gl_obshij_podhod}, leading to the correction formula similar to (\ref{eq_delta_g_c}). The results of Kuroda \& Mio for EST and ESD schemas were confirmed as special cases of the formula (\ref{eq_delta_g_c}). Though the paper \cite{nagornyi1995} contained all necessary information about the coefficient $C_1$ to write down the formula (\ref{eq_c_corr_ESD_V0}), no further analysis of the correction was done at the paper.
\subsection{Robertsson
\cite{robertsson2005}
}
Using the signal delays approach, the distortion component of the trajectory was found as in formula (\ref{eq_c_z_reg_path}). Projecting the distortion component to the space spanned by the shifted Legendre basis, the correction for the case of a test mass falling away from the beam splitter was found as
\begin{equation}
\label{eq_c_robertsson}
\Delta g_c =  \frac{3 g_0 V_0}{c} +
\frac{3 T g_0^2 \left( 1 + \eta_2(\lambda)/5 \right)}{2c},
\end{equation}
where
\begin{equation}
\label{eq_c_robertsson_eta}
\eta_2(\lambda) = \frac{5 \lambda (\lambda^2-3)}{7 (3 \lambda^2-5)},
\end{equation}
where
\begin{equation}
\label{eq_c_robertsson_lambda}
\lambda=\cases
{
\frac{1}{1+2 V_0 / g_0 T} &, for ESD schemas,\\
0 &  , for EST schemas.\\
}
\end{equation}
If put together, the above equations yield the same correction formulas as (\ref{eq_c_corr_eq_T}) and (\ref{eq_c_corr_ESD_V0}).
\section{Conclusions}
In the paper we attempted to revise the correction due to finite speed of light in absolute gravimeters in order to understand and reconcile differences existing in the theory. Like other authors, we based our reasoning on two interrelated physical phenomena: delay in the electromagnetic wave propagation, and the Doppler effect. To achieve our goal, we had to make the following advancements in the implementation of the phenomena models:
            \begin{enumerate}
            \item Introduction of two scales of time associated with moving and resting reflectors and deriving the set of rules allowing for easy transition between the scales;
            \item Establishing the equivalence between the observed trajectory distortions derived as either delays in registered time, or progressions of registered distance.
            \end{enumerate}
The analysis we made leads to the following conclusions:
\begin{enumerate}
\item The Doppler effect in the scale of time associated with the test body is responsible for the $\frac{2}{3}$ of the correction due to finite speed of light in absolute gravimeters. Transition of the test body's velocity to the time scale associated with the beam splitter, which is equivalent to introduction of optical signal delays, is necessary to get the full correction value..
\item For EST schemas implementing least-squares fitting of the three-parameter linear model (working formula (\ref{eq_c_G3})), the correction due to finite speed of light does not depend on the number of levels and always defined by the formula (\ref{eq_c_corr_eq_T}).
\item For ESD schemas implementing the same model, the correction is a convoluted function of number of levels and initial velocity. Approximation by Kuroda \& Mio (formula (\ref{eq_c_corr_ESD_V0_0})) may be up to 1 $\mu$Gal off for real instruments. The approximation by formula (\ref{eq_c_corr_ESD_V0}) is 0.01 $\mu$Gal accurate, if the number of levels is at least several hundreds.
\item If the finite speed of light is accounted for by correcting the measured time intervals, the parameter ``$z_0$'' in time recalculations can be dropped for the following schemas:
    \begin{itemize}
    \item three-level schema (\ref{eq_c_g_izm_G_3_level}),
    \item four-level schema (\ref{eq_c_g_izm_G_4_level}),
    \item any multi-level schema implementing least-squares fitting of the three-parameter linear model (\ref{eq_c_G3}).
    \end{itemize}
\item Discrepancies in the corrections obtained by different authors are caused by several reasons, including confusion between corrections due to the Doppler effect and due to finite speed of light, use of simplified trajectory models, and math errors.
\end{enumerate}
\appendix
\setcounter{section}{0}
\section{Weighting functions of absolute gravimeters}
\subsection{Working formulas of absolute gravimeters}
\begin{figure}[ht]
\centering
\small
\input{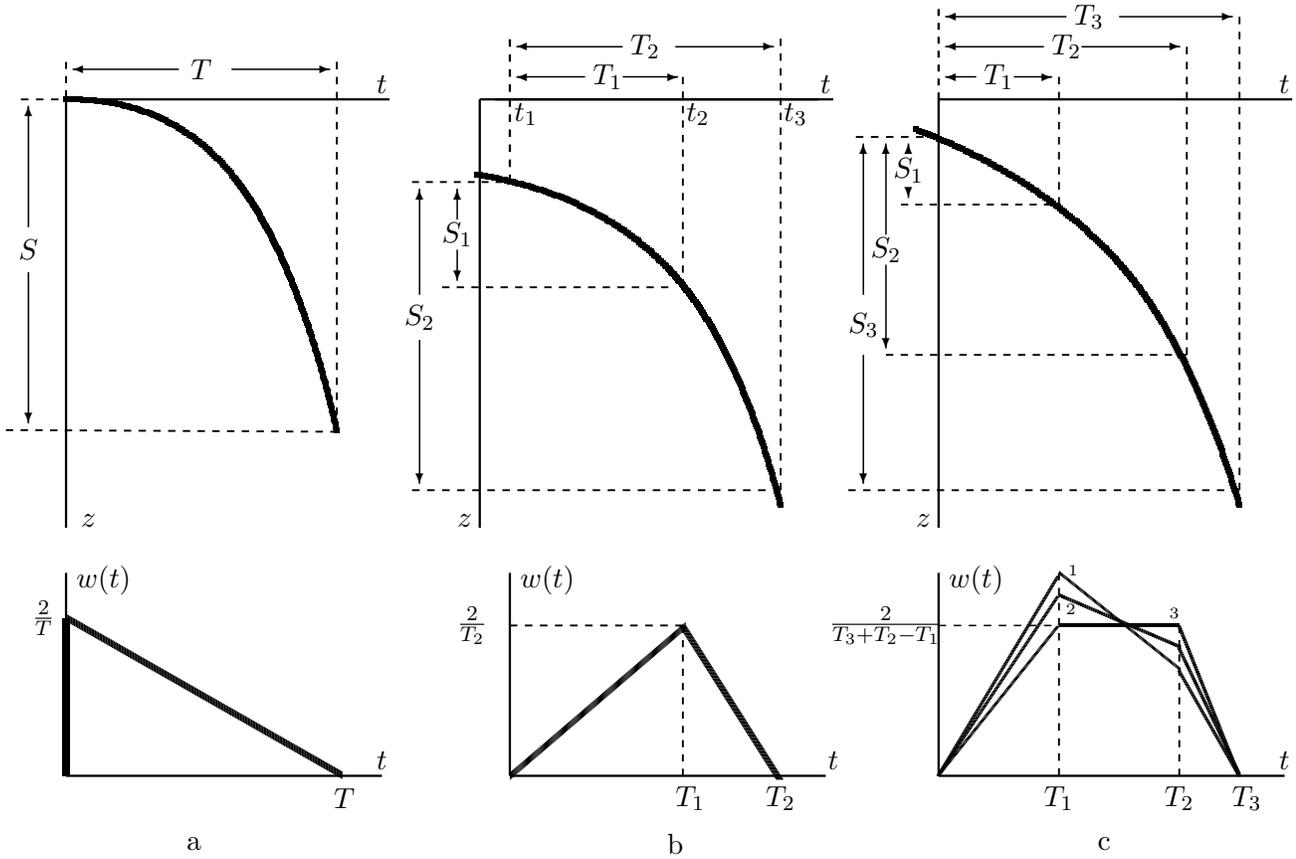}
  \caption[short title]
  {
  \quad\parbox[t]{13cm} {Measurement schemas of absolute gravimeters and corresponding weighting functions:
      \begin{compactdesc}
      \item[a: ]{two-level schema;}
      \item[b: ]{three-level schema;}
      \item[c: ]{four-level schema, the weighting function depends on the working formula:\\
      1 -- working formula (\ref{eq_c_G2}),\\
      2 -- working formula (\ref{eq_c_G3}),\\
      3 -- working formula (\ref{eq_c_g_izm_G_4_level}).
      }
      \end{compactdesc}
    }
  }
\label{pic: 2-3-4_levels}
\end{figure}

Weighting function of absolute gravimeters are based on their working formulas that map measured from the common start intervals of time $\{T_i\}_{i=1}^N$ and distance $\{S_i\}_{i=1}^N$ to the measured gravity acceleration\footnote{Because intervals are measured between levels, the number of levels is always one more that the number of intervals.}:
\begin{equation}
\label{eq_c_g_izm_G}
\overline g = G \left( T_i \, , \, S_i \right).
\end{equation}
The term ``working formula'' we use here is narrower than the term ``measurement equation'', which usually includes all the calculations leading to the measurement result \cite{kacker2007}.
It's essential that working formulas are linear in measured distances. In the paper, the following working formulas are considered:
\begin{itemize}
\item Two-level schema (fig.~\ref{pic: 2-3-4_levels}a, $N=1$)
\begin{equation}
\label{eq_c_g_izm_G_2_level}
\overline g = G \left( T \, , \, S \right) = \frac{2\,S}{T^2}.
\end{equation}
\item Three-level schema (fig.~\ref{pic: 2-3-4_levels}b, $N=2$)
\begin{equation}
\label{eq_c_g_izm_G_3_level}
\overline g = G \left( T_i \, , \, S_i \right) =
\left( \frac{S_2}{T_2} - \frac{S_1}{T_1} \right) \frac{2}{T_2-T_1}.
\end{equation}
\item Four-level schema (fig.~\ref{pic: 2-3-4_levels}c, $N=3$)
\begin{equation}
\label{eq_c_g_izm_G_4_level}
\overline g = G \left( T_i \, , \, S_i \right) =
\left( \frac{S_3-S_2}{T_3-T_2} - \frac{S_1}{T_1} \right) \frac{2}{T_3+T_2-T_1}.
\end{equation}
\item Multi-level schema with least-squares fitting of two-parameter model $S_i = V_0t+g_0 t^2/2$
\begin{equation}
\label{eq_c_G2}
\overline g = G \left( T_i \, , \, S_i \right)
=2\left|
\begin{array}{cc}
 \sum T_{i} ^{2} & \sum T_{i} S_{i} \\
 \sum T_{i}^{3} & \sum T_{i}^{2} S_{i}
\end{array}
\right| :\left|
\begin{array}{cc}
\sum T_{i} ^{2} & \sum T_{i} ^{3} \\
\sum T_{i}^{3} & \sum T_{i}^{4}
\end{array}
\right|.
\end{equation}
\item Multi-level schema with least-squares fitting of three-parameter model $S_i = z_0+V_0t+g_0 t^2/2$
\begin{equation}
\label{eq_c_G3}
\overline g = G \left( T_i \, , \, S_i \right)
=2\left|
\begin{array}{ccc}
N & \sum T_{i} & \sum S_{i} \\
\sum T_{i} & \sum T_{i} ^{2} & \sum T_{i} S_{i} \\
\sum T_{i}^{2} & \sum T_{i}^{3} & \sum T_{i}^{2} S_{i}
\end{array}
\right| :\left|
\begin{array}{ccc}
N & \sum T_{i} & \sum T_{i} ^{2} \\
\sum T_{i} & \sum T_{i} ^{2} & \sum T_{i} ^{3} \\
\sum T_{i}^{2} & \sum T_{i}^{3} & \sum T_{i}^{4}
\end{array}
\right|.
\end{equation}
\end{itemize}
\subsection{Determination of weighting functions}
\begin{figure}[h]
\centering
\small
\input{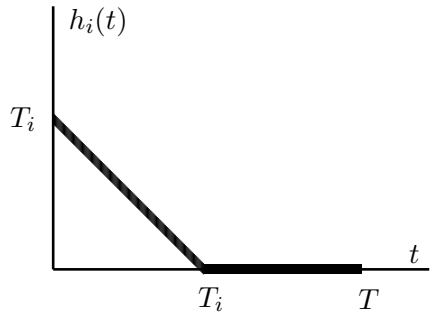}
  \caption[short title]
  {
  \quad\parbox[t]{11cm} {The function $h_i(t)$ (\ref{eq_c_h_i}) that generates gravimeter's weighting function, when substituted for the $S_i$ interval to the gravimeter's working formula $G \left( T_i \, , \, S_i \right)$.
  }
  }
\label{pic_c_hi}
\end{figure}

\begin{figure}[ht]
\centering
\small
\input{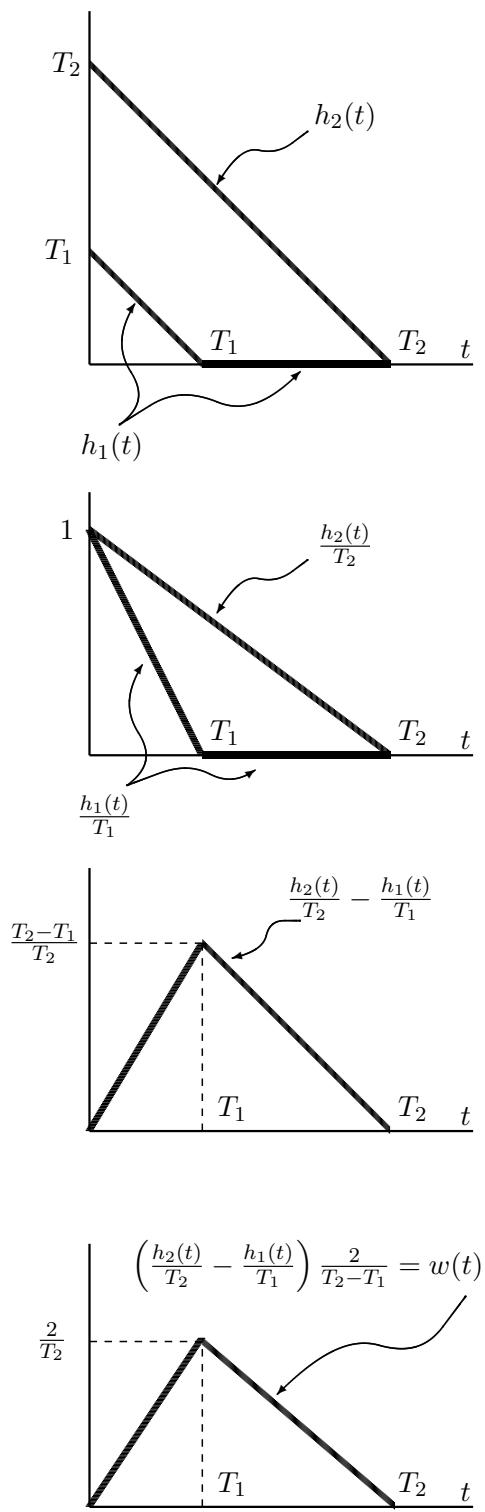}
  \caption[short title]
  {
  \quad\parbox[t]{11cm} {Formation of the weighting function for the three-level schema
  }
  }
\label{pic_c_1}
\end{figure}

\begin{figure}[ht]
\centering
\small
\input{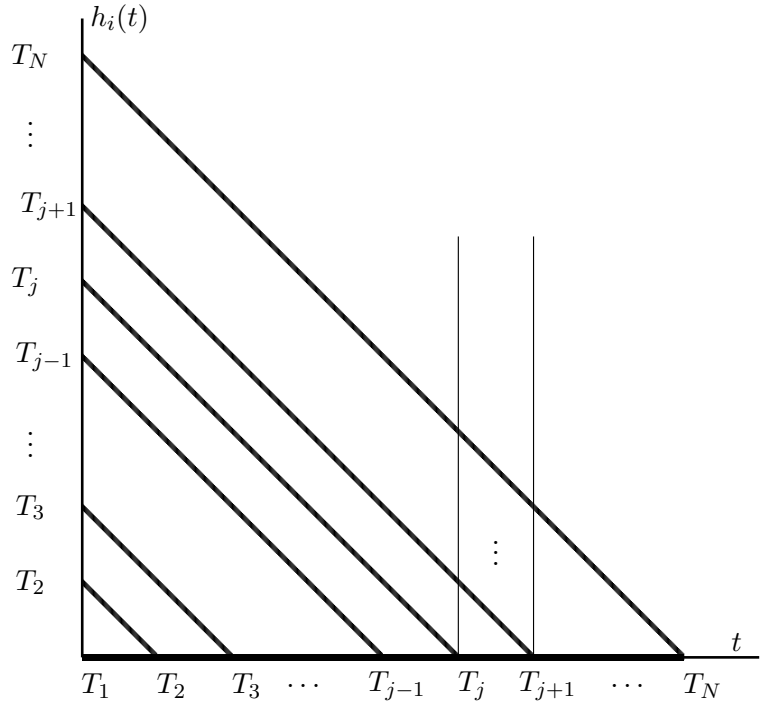}
  \caption[short title]
  {
  \quad\parbox[t]{11cm} {For the formula (\ref{eq_c_Sij}): on the time interval $T_{j}$\ldots$T_{j+1}$ only those $h_i(t)$ components are non-zero, for which $i>j$.
  }
  }
\label{pic_c_2}
\end{figure}

\begin{figure}[ht]
%
\PSforPDF{
%
%
%
\newcommand{\myfbox}{\fbox}
\begin{psfrags}
\psfrag{x05}[tc][tc]{$0.05$}
\psfrag{x10}[tc][tc]{$0.10$}
\psfrag{x15}[tc][tc]{$0.15$}
\psfrag{x20}[tc][tc]{$0.20$}
\psfrag{x25}[tc][tc]{$0.25$}
\psfrag{x30}[tc][tc]{$0.30$}
\psfrag{y5}[tc][tc]{$5$}
\psfrag{y10}[tl][tc]{$10$}
\psfrag{y15}[tc][tc]{$15$}
\psfrag{y20}[tc][tc]{$20$}
\psfrag{t}[tc][tc]{$t$, s}
\psfrag{w}[tl][tc]{$w(t)$, s$^{-1}$}
%
%
%
%
%
%
\myfbox{\includegraphics[width=16cm, trim=0 -5 -5 -10]{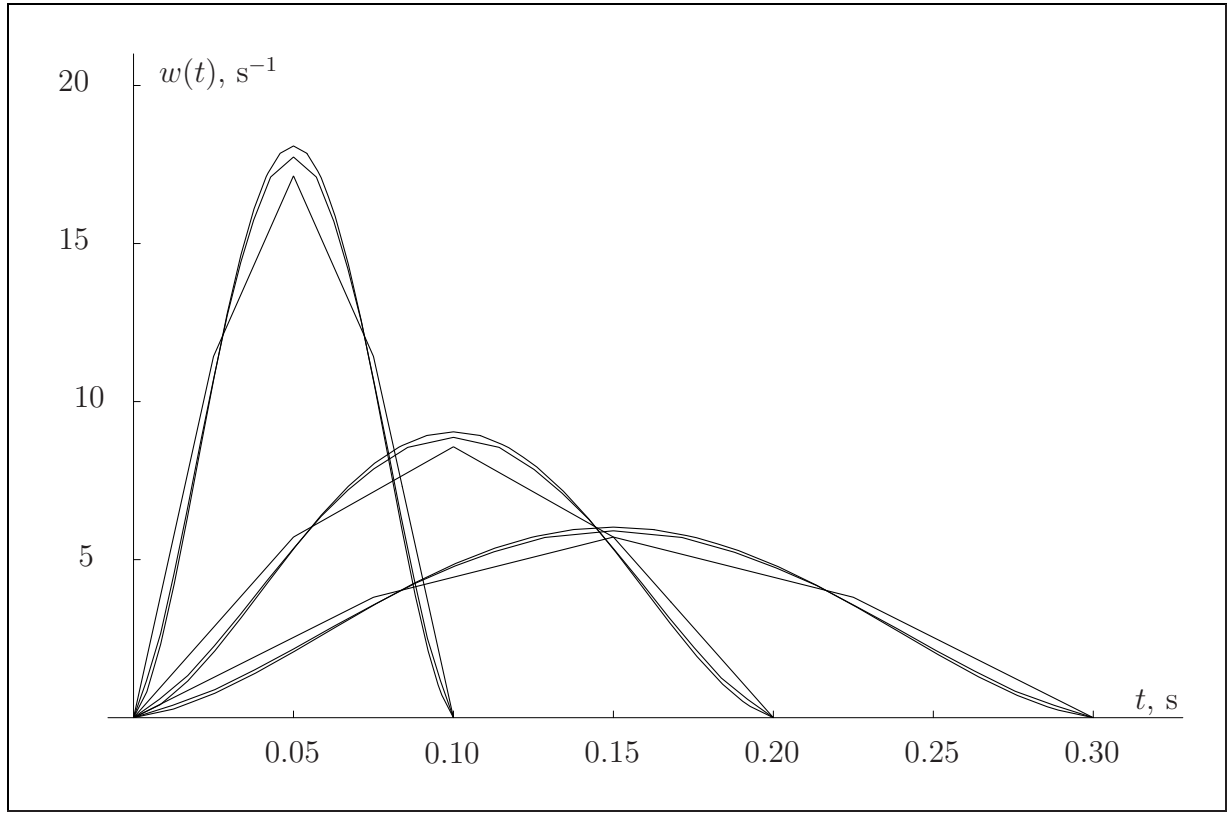}}
\end{psfrags}
}
  \caption[short title]
  {
  \quad\parbox[t]{10cm} {Weighting functions for the formula (\ref{eq_c_G3}) with measurement intervals of 0.1, 0.2 и 0.3 s. The EST levels are 5, 15, and 25.}
  }
\label{pic_c_wfs.eps}
\end{figure}

To find the weighting function of a gravimeter with working formula $G \left( T_i \, , \, S_i \right)$ linear in distance intervals $S_i$, one should replace $S_i$ with piecewise linear functions $h_i(t)$ defined as \cite{nagornyi1995}(fig.~\ref{pic_c_hi}) :
\begin{equation}
\label{eq_c_h_i}
h_i(t)=\cases
{
T_i - t & $0\le t \le T_i$,\\
0 &  $T_i\le t \le T$\\
}
\end{equation}
that means
\begin{equation}
\label{eq_c_w_via_h}
w(t) = G \left( T_i \, , \, h_i(t) \right).
\end{equation}
Weighting function for the two-level schema (\ref{eq_c_g_izm_G_2_level}), will therefore be generated by a single component like (\ref{eq_c_h_i}):
\begin{equation}
\label{eq_c_w_l_level}
w(t) = 2 \, \frac{T-t}{T^2}.
\end{equation}
This weighting function is shown on the fig.~\ref{pic: 2-3-4_levels}a.
For the three-level schema (\ref{eq_c_g_izm_G_2_level}), the weighting function is generated, according to its working formula, by a linear combination of two scaled components like  (\ref{eq_c_h_i}). This process is shown in detail on the fig.\ref{pic_c_1}
and results in the triangle-shaped weighting function (fig.~\ref{pic: 2-3-4_levels}b):
\newpage
\begin{equation}
\label{eq_c_w_2_level}
w(t)=\cases
{
\frac{2\,t}{T_1 \, T_2} & $0\le t \le T_1$, \\
\\
\frac{2\,(T_2 - t)}{T_2\,(T_2 - T_1)} & $T_1 \le t \le T_2$ .
}
\end{equation}
For the four-level schema (\ref{eq_c_g_izm_G_4_level}) the weighing function is trapeze-shaped (fig.~\ref{pic: 2-3-4_levels}c-3):
\newpage
\begin{equation}
\label{kdu_1.58}
w(t)=\cases
{
\frac{2t}{T_1(T_3+T_2-T_1)} & $0 \le t \le T_1$,
\\
\frac{2}{T_3+T_2-T_1} &$T_1 \le t \le T_2$,
\\
\frac{2(T_3-t)}{(T_3-T_2)(T_3+T_2-T_1)} &$T_2 \le t \le T_3$.
}
\end{equation}
In the same way, weighting functions of multi-level schemas are linear combinations of $N$ components like (\ref{eq_c_h_i}), resulting in piecewise linear weighting functions with $N-1$ segments. As shown on the fig.~\ref{pic_c_2},
the $j$-th segment of the weighting function is created only by those $h_i(t)$ components, for which $i>j$. Than is,
\begin{equation}
\label{eq_c_wj}
w_j(t)=G \left( T_i\;,\;S_{ji} \right),
\end{equation}
where
\begin{equation}
\label{eq_c_Sij}
S_{ij}=\left\{
\begin{array}{ccl}
0 & , & i= \overline{1\,,\,j}, \\
T_i-t & , & i=\overline{j+1\,,\,N}.
\end{array}
\right.
\end{equation}
Weighting functions obtained with the formulas (\ref{eq_c_wj}) and (\ref{eq_c_Sij}) are shown on the fig.~\ref{pic_c_wfs.eps}.
%
%
%
%
%
%
As the number of levels increases, the functions converge to a  limiting shape. For the EST schema  with working formula (\ref{eq_c_G3}), the limiting shape is defined by
 \begin{equation}
\label{eq_c_wf_T_lim}
w(t) = 30\frac{t^2}{T^3} - 60\frac{t^3}{T^4} + 30\frac{t^4}{T^5}.
\end{equation}
For the ESD schemas with working formula (\ref{eq_c_G3}), the limiting shape depends on the initial velocity and is defined by the equation:
\begin{eqnarray}
\label{w3s_lim}
w(t)= &
\left( \frac{60\,t^4}
       {T^5} -
      \frac{120\,t^3}{T^4} +
      \frac{60\,t^2}{T^3}
      \right)
\\
& \times \frac{
 g_0^3\,t\,T^2 +
      3\,g_0^2\,T\,
       \left( 2\,t + T \right)
         \,V_0 +
      6\,g_0\,
       \left( t + 2\,T \right)
         \,V_0^2 + 10\,V_0^3
 }{g_0^3\,T^3 +
    12\,g_0^2\,T^2\,V_0 +
    30\,g_0\,T\,V_0^2 + 20\,V_0^3}
     \nonumber
.
\end{eqnarray}
%
%
%
%
%
\begin{figure}[ht]
%
\PSforPDF{
%
%
%
\newcommand{\myfbox}{\fbox}
\begin{psfrags}
\psfrag{x0}[tc][tc]{$0$}
\psfrag{x025}[tc][tc]{$0.25$}
\psfrag{x05}[tc][tc]{$0.5$}
\psfrag{x075}[tc][tc]{$0.75$}
\psfrag{x01}[tc][tc]{$1$}
\psfrag{y5}[tc][tc]{$5$}
\psfrag{y10}[tl][tc]{$10$}
\psfrag{y15}[tc][tc]{$15$}
\psfrag{y20}[tc][tc]{$20$}
\psfrag{t}[tr][tc]{$\;t \times 10 $, s}
\psfrag{w}[tl][tc]{$w(t)$, s$^{-1}$}
\psfrag{one}[tl][tc]{$V_0=0$}
\psfrag{two}[tl][tc]{$V_0=0.2$ m/s}
\psfrag{three}[tl][tc]{$V_0 \rightarrow \infty$}
%
%
%
%
%
%
\myfbox{\includegraphics[width=16cm, trim=0 -5 -5 -10]{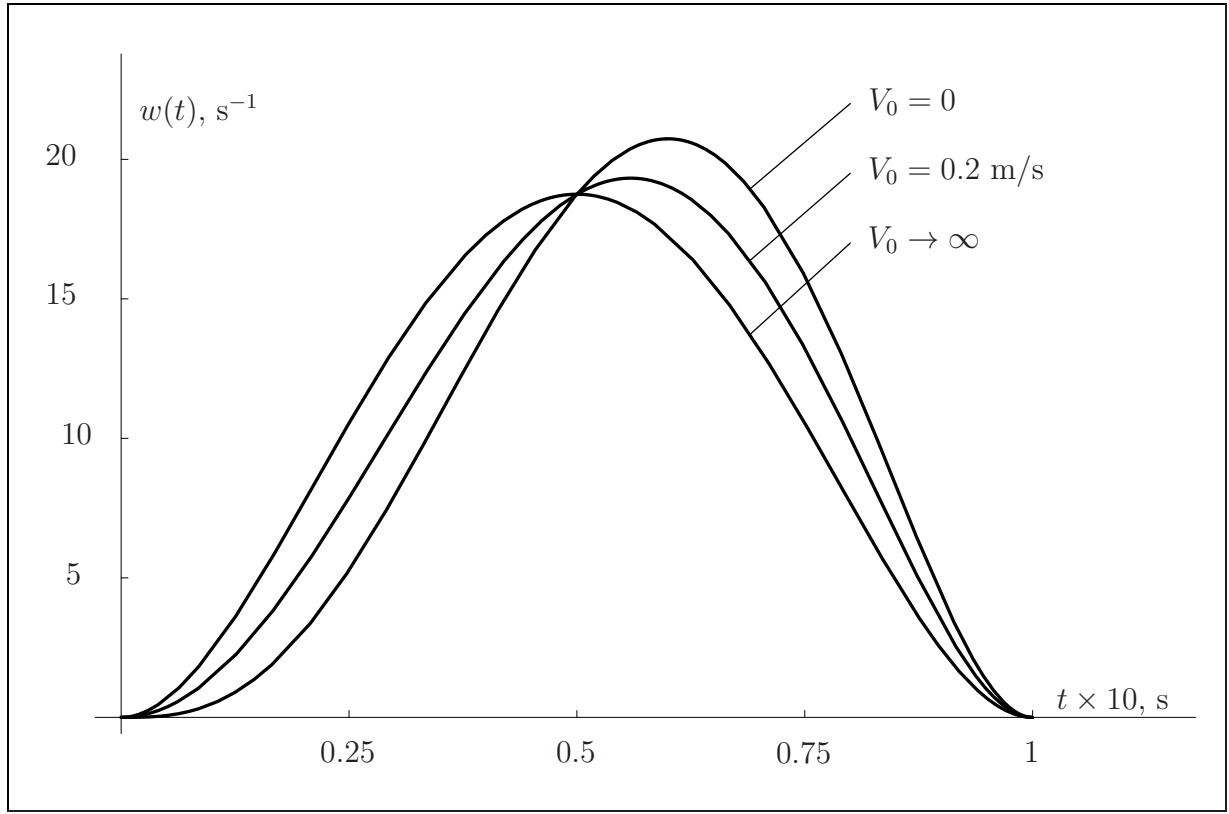}}
\end{psfrags}
}
  \caption[short title]
  {
  \quad\parbox[t]{10cm} {Limiting shapes for the WFs of the ESD schemas for different initial velocities (\ref{w3s_lim}). $g_0=9.81$ m/s$^{2}$}
  }
\label{pic_c_w3s_V0}
\end{figure}
%
%
%
Weighting functions for the ESD schema approach those for the EST schema as $V_0 \rightarrow \infty$, but even for small initial velocities they are significantly deviating from the case $V_0=0$ (fig.~\ref{pic_c_w3s_V0}).
A method of finding limiting shapes can be found in \cite{nagornyi1995}.

The working formulas (\ref{eq_c_G2}) and (\ref{eq_c_G3}) can also be applied to the schemas with three and four levels. For three levels, the weighting function is still the same triangle (fig.~\ref{pic: 2-3-4_levels}b), as there is only one parabola passing through three points. For four levels, the weighting function significantly depends on the working formula (fig.~\ref{pic: 2-3-4_levels}c).
\section{Averaging coefficients of absolute gravimeters}
Based on the generic formula for weighting functions (\ref{eq_c_w_via_h}), the averaging coefficients can be found as \cite{nagornyi1995}
\begin{equation}
\label{eq_c_Cn}
C_n = \frac{G\left(T_i \, , \, T_i^{n+2}\right)}{(n+1)(n+2)}.
\end{equation}
For the correction due to finite speed of light, only the $C_1$ coefficient is needed.
According to (\ref{eq_c_Cn}), this coefficient is
\begin{itemize}
\item For the two-level schema (\ref{eq_c_g_izm_G_2_level}):
\begin{equation}
\label{eq_c_C1_2_level}
C_1 = \frac{T}{3}.
\end{equation}
\item For the three-level schema (\ref{eq_c_g_izm_G_3_level}):
\begin{equation}
\label{eq_c_C1_3_level}
C_1 = \frac{T_1 + T_2}{3}.
\end{equation}
\item For the four level schema (\ref{eq_c_g_izm_G_4_level}):
\begin{equation}
\label{eq_c_C1_4_level}
C_1 = \frac{T_3^2 + T_2^2 - T_1^2 + T_2 T_3}{3(T_3+T_2-T_1)}.
\end{equation}
\item For the schema with four levels and working formula  (\ref{eq_c_G2}):
\begin{equation}
\scriptstyle
\label{eq_c_C1_4_multilevel}
C_1 = \frac{1}{3}
\frac
{
T_1^2 T_2^2(T_1 - T_2)^2(T_1 + T_2) +
T_1^2 T_3^2(T_1 - T_3)^2(T_1 + T_3) +
T_2^2 T_3^2(T_2 - T_3)^2(T_2 + T_3)
}
{
T_1^2 T_2^2(T_1 - T_2)^2 +
T_1^2 T_3^2(T_1 - T_3)^2 +
T_2^2 T_3^2(T_2 - T_3)^2
}
.
\end{equation}
%
\item EST schemas. If levels equally spaced in time, the $i$-th time interval is
\begin{equation}
\label{eq_Ti_timed}
T_i = T\,\frac{i-1}{N-1},
\end{equation}
where $T$ is the duration of the measurement interval. Substitution of (\ref{eq_Ti_timed}) into (\ref{eq_c_Cn}) creates to the following sums:
\begin{equation}
\label{eq_Summa_n_Ti}
\sum \limits_{i=1}^N \left( \frac{i-1}{N-1} \right)^n.
\end{equation}
For different $n$'a, the values of the sums are available in references \cite{abramowitz1964}.
Substituted to the formulas for $C_1$
(\ref{eq_c_Cn},~\ref{eq_c_G3}), the values result in
\begin{equation}
\label{eq_C1_T}
C_1 = \frac{T}{2}.
\end{equation}
Another way to prove this fact would be to observe, that regardless of the number of the levels, weighting functions for the EST schemas are always symmetric with regard to the middle point (fig.~\ref{pic_c_wfs.eps}), so
\begin{equation}
\label{c_wf_sym}
\int \limits_{0}^{T} t\;w(t)\;\D t =
\int \limits_{0}^{T} \left(t-\frac{T}{2}\right)\;w(t)\;\D t +
\frac{T}{2}  \int \limits_{0}^{T} w(t)\;\D t =
\frac{T}{2}.
\end{equation}
The first integral equals to 0 because of the symmetry, the second integral equals 1 sue to the unit property (\ref{eq_wf_unit_square}).
\item ESD schemas.
If $N$ levels are equally spaced along the total path $S$, the $i$-th level can be found as
\begin{equation}
\label{eq_Si_spaced}
S_i = S\,\frac{i-1}{N-1}.
\end{equation}
The time interval, corresponding to the $S_i$ can be found using
\begin{equation}
\label{eq_Si_no_V0}
S_i = V_0 T_i + g_0 T_i^2/2.
\end{equation}
This equations doesn't include initial displacement, because
for calculations the origin of the time axis can be co-located with the first measured level\footnote{This shift of the time origin, however, does not equate the models $S_i = V_0 T_i + g_0 T_i^2/2$ and $S_i = z_0+ V_0 T_i + g_0 T_i^2/2$  when used for the least-squares fitting of the trajectory. These models have different weighting functions in terms of (\ref{eq_c_g_izm}) and so they average the disturbances in the test body motion differently \cite{nagornyi1995}.}.
Using the length of the trajectory
\begin{equation}
\label{eq_T}
S = V_0 T + g_0 T^2/2,
\end{equation}
where $T=T_N$ is the total measurement time,
we get
\begin{equation}
\label{eq_Ti_spaced}
T_i = \frac{-V_0+ \sqrt{V_0^2 + 2g_0\left(V_0T + g_0 T^2/2\right)\frac{i-1}{N-1}}}{g_0}.
\end{equation}
This $T_i$ vector we need to substitute into the equation (\ref{eq_c_Cn}), which for $n=1$ becomes
\begin{equation}
\label{eq_c_C1}
C_1 = \frac16 \; G\left(T_i \, , \, T_i^3\right),
\end{equation}
to obtain the $C_1$ coefficient, which turns out to be a complicated function of $g_0$, $V_0$, $N$, and $T$.
If $N\to \infty$, the $C_1$ can be found using the limit weighting function. Substituting (\ref{w3s_lim}) into (\ref{eq_c_Cn_as_Int}), we get
\begin{equation}
\label{c_ESD_C1_VO}
C_1 = \frac{T\,\left( 4\,g_0^3\,T^3 +
      45\,g_0^2\,T^2\,V_0 +
      108\,g_0\,T\,V_0^2 + 70\,V_0^3
      \right) }{7\,
    \left( g_0^3\,T^3 +
      12\,g_0^2\,T^2\,V_0 +
      30\,g_0\,T\,V_0^2 + 20\,V_0^3
      \right) }.
\end{equation}
For the case $V_0=0$, the above expression simplifies to
\begin{equation}
\label{c_ESD_C1_VO_0}
C_1 = \frac47 \; T.
\end{equation}
In the chapter~\ref{c_po_puti}  we investigate
how accurate the two above formula are when applied to real instruments.
\item For the symmetric rise-and-fall measurement schemas
\begin{equation}
\label{eq_c_C1_symm}
C_1 = 0,
\end{equation}
which follows from the evenness of weighting functions for such schemas \cite{nagornyi1995}.
\end{itemize}
\section*{References}

\bibliographystyle{ieeetr}
\end{document}